\documentclass[prd,preprint,nofootinbib,superscriptaddress]{revtex4}
\usepackage{cancel}
\usepackage[dvipdfm]{graphicx}
\usepackage{amsmath,amssymb}
\begin{document}
\newcommand{\gsim}{ \mathop{}_{\textstyle \sim}^{\textstyle >} }
\newcommand{\lsim}{ \mathop{}_{\textstyle \sim}^{\textstyle <} }
%%%%%%%%%%%%%%%%%%%%%%%%%%%%%%%%%
\renewcommand{\thefootnote}{\fnsymbol{footnote}}

\vspace*{1cm}
\title{An R-parity conserving radiative neutrino mass model \\
without right-handed neutrinos}
\renewcommand{\thefootnote}{\alph{footnote}}
\preprint{KANAZAWA-10-03}
\preprint{UT-HET 038}
\preprint{KU-PH 005}
\author{Mayumi Aoki\footnote{mayumi@hep.s.kanazawa-u.ac.jp}}
\affiliation{Institute for Theoretical Physics, Kanazawa University, Kanazawa 920-1192, Japan}
\author{Shinya Kanemura\footnote{kanemu@sci.u-toyama.ac.jp}}
\affiliation{Department of Physics, University of Toyama, Toyama 930-8555, Japan}
\author{Tetsuo Shindou\footnote{shindou@cc.kogakuin.ac.jp}}
\affiliation{Department of Technology, Kogakuin University, Tokyo 163-8677, Japan}
\author{Kei Yagyu\footnote{keiyagyu@jodo.sci.u-toyama.ac.jp}}
\affiliation{Department of Physics, University of Toyama, Toyama 930-8555, Japan}
\begin{abstract}
\noindent
The model proposed by A.~Zee (1986) and K.~S.~Babu (1988) is a 
 simple radiative seesaw model, in which tiny neutrino masses are 
generated at the two-loop level. We investigate a supersymmetric extension of 
the Zee-Babu model under R-parity conservation. 
The lightest superpartner particle can then be a dark matter candidate.
We find that the neutrino data can be reproduced with satisfying 
current data from lepton flavour violation even in the scenario where
not all the superpartner particles are heavy.
Phenomenology at the Large Hadron Collider is also discussed.
\end{abstract}

\maketitle

%\section{Introduction}
%It is important to consider a new physics model at TeV scale.
Although the standard model (SM) has been successful in describing phenomena 
below 100\;GeV, the Higgs sector has not been confirmed yet.
The Higgs boson is expected to be lighter than one TeV from the
unitarity argument, 
so that it can be explored at the CERN Large Hadron Collider (LHC).
On the other hand, we require new physics beyond the SM because of several reasons,
such as the quadratic divergence problem,
the origin of tiny masses of neutrinos,  
the existence of dark matter (DM), and so on.
It is interesting to consider a scenario where these problems are
simultaneously solved at the TeV scale, as such a scenario is directly
testable at the LHC or future colliders such as the International Linear
Collider (ILC) and the Compact Linear Collider (CLIC). 
In such a case, it is plausible that the Higgs sector is closely related
to the detail of physics beyond the SM.

One of the important motivations to consider physics beyond the SM is 
to explain the origin of tiny neutrino masses.
The seesaw mechanism is known to be a simple method of generating neutrino
masses at the tree level, in which tiny masses of the (left-handed)
neutrinos may be obtained from 
very heavy right-handed neutrinos (type I)\cite{type1}, 
a heavy triplet Higgs boson (type II)\cite{type2}, 
or a heavy triplet fermion (type III)\cite{type3}.
However,  
the mass scale of these fields is much higher than the TeV scale;
naively at around $\mathcal{O}(10^{6-15})\;\text{GeV}$,
unless the coupling constants between lepton doublets and
these new heavy fields are taken to be unnaturally small. 
Such a high scale is far from experimental reach.

Radiative seesaw models, where neutrino masses are generated at the
quantum corrections, are alternative attractive scenarios to generate 
tiny neutrino masses\cite{zee,zee-2loop,babu-2loop,Krauss:2002px,Ma:2006km,Aoki:2008av}. 
Masses of new particles in these models can
be as low as the TeV scale,
so that they are expected to be directly testable at current and future
collider experiments.
One of the characteristic features of these models is an extended Higgs sector.
Another feature is the Majorana nature, either introducing lepton 
number violating couplings or introducing right-handed neutrinos.

The original model for radiative neutrino mass generation was first proposed 
by A.~Zee\cite{zee}, where neutrino masses are generated at the 
one-loop level by adding an extra  $\text{SU}(2)_{\text{L}}^{}$ doublet scalar field
and a charged singlet scalar field with lepton number violating
couplings to the SM particle entries.
Phenomenology of this model has been studied in Ref.~\cite{zee-ph1}.
However, it turned out that it was difficult to reproduce the current
data for neutrino oscillation in this original model\cite{zee-nu}.
Some extensions have been discussed in Ref.~\cite{zee-variation}.

The simplest successful model today may be the one proposed by
A.~Zee\cite{zee-2loop} and K.~S.~Babu\cite{babu-2loop},
in which two kinds of $\text{SU}(2)_{\text{L}}^{}$ singlet scalar fields
are introduced; i.e., a singly charged scalar boson and a
doubly charged one.
These fields carry lepton number of two unit. 
In this model, which we refer to as the Zee-Babu model, 
the neutrino masses are generated at the two-loop level.
Phenomenology of this model has been discussed in
Refs.~\cite{macesanu,aristizabal,nebot,ohlsson,ak-letter}.
Apart from the Zee-Babu model, there is also another type of radiative seesaw
models\cite{Krauss:2002px,Ma:2006km,Aoki:2008av}, where 
TeV-scale right-handed neutrinos are introduced with the odd charge
under the discrete $Z_2^{}$ symmetry. 
In these models, the $Z_2^{}$ symmetry protects the decay of the lightest
$Z_2^{}$ odd particle, which can be a candidate of DM.
This is an advantage of this class of models\cite{knt_dm,ma_dm,aks_dm}.
On the other hand, in the Zee-Babu model there is no such a discrete
symmetry and no neutral new particle, so that there is no DM candidate.

In this Letter, we investigate a supersymmetric extension 
of the Zee-Babu model.  
By introducing supersymmetry (SUSY), the quadratic divergence
in the one-loop correction to the mass of the Higgs boson  
can be eliminated automatically.
In addition, a discrete symmetry, which is so called the R-parity, is
imposed in our model to forbid the term which causes the dangerous proton decay.
The R-parity also guarantees the stability of the lightest super partner
particle (LSP) such as the neutralino,
which may be identified as a candidate of DM.
We find that there are allowed parameter regions in which 
the current neutrino oscillation data can be reproduced
 under the constraint from the lepton flavour
violation (LFV) data. 
In addition, this model provides quite interesting phenomenological signals
in the collider physics; i.e., the existence of singly as well as
doubly charged singlet scalar bosons and their SUSY partner fermions.
Such an allowed parameter region also appears even when new particles and their partners
are as light as the electroweak scale. 
We also discuss the outline of phenomenology for these particles at
the LHC.

\begin{figure}
\begin{center}
{\includegraphics[scale=0.8]{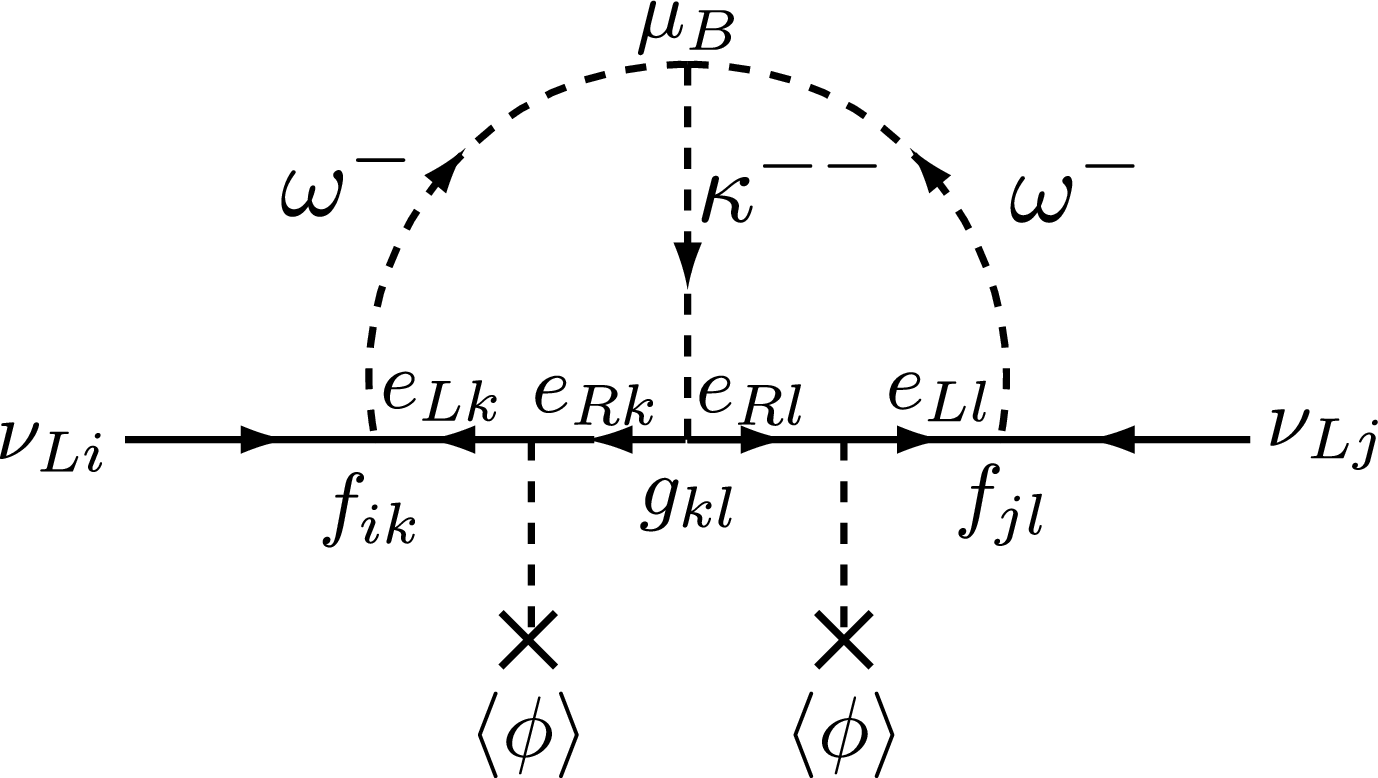}}
\end{center}
\caption{The two-loop diagram relevant to the neutrino mass matrix.}
\label{fig:diagram0}
\end{figure}

%\section{Model}
In the original (non-SUSY) Zee-Babu model\cite{zee-2loop,babu-2loop},
two kinds of $\text{SU}(2)_{\text{L}}^{}$ singlet fields $\omega^{-}$ ($Y=-1$) and
$\kappa^{--}$ ($Y=-2$) are introduced.
The Yukawa interaction and the scalar potential are given by 
\begin{equation}
\mathcal{L}=
-\sum_{i,j=1}^3 f_{ij}^{}\bar{\ell}_{Li}^{c}\cdot \ell_{Lj}^{}\omega^+
-\sum_{i,j=1}^3g_{ij}^{}\bar{e}_{Ri}^{}e_{Rj}^c\kappa^{--}
-\mu_B^{}\omega^{-}\omega^{-}\kappa^{++}+\text{h.c.}-V^{\prime}-V_{\text{SM}}^{}\;,
\end{equation}
where $V_{\text{SM}}^{}$ is the Higgs potential of the SM,  
$\ell_{Li}^{}$ are lepton doublets, $e_{Ri}^{}$ are right-handed charged leptons,
the indices $i$, $j$ are the flavour indices, 
the dot product of the fields denotes 
the antisymmetric contract of the $\text{SU}(2)_{\text{L}}^{}$ indices i.e. 
$\bar{\ell}_{Li}^c\cdot \ell_{Lj}^{}\equiv 
\sum_{\alpha,\beta=1}^2
\epsilon_{\alpha\beta}^{}\bar{\ell}_{Li}^{\alpha c}\ell_{Lj}^{\beta}$, and all the scalar couplings with
respect to $\omega^-$ and $\kappa^{--}$ other than 
$\omega^-\omega^-\kappa^{++}$ are in $V^{\prime}$.
Notice that lepton number conservation is broken only by the term of $\mu_B ^{}$.
The neutrino mass matrix is generated via two-loop diagrams as shown in Fig.~\ref{fig:diagram0}.
The induced neutrino mass matrix is computed as\footnote{Our result for
the neutrino mass matrix is consistent with that in Ref.~\cite{nebot}
including the factor.} 
\begin{equation}
(m_{\nu}^{})_{ij}^{}=\left(\frac{1}{16\pi^2}\right)^2\sum_{k,l=1}^3
\frac{16\mu_B^{} f_{ik}^{}(m_{e}^{})_k^{} g_{kl}^{}(m_e^{})_{l}^{}f_{jl}^{}}{m_{\kappa}^2}
 I(m_{\omega}^{},(m_e^{})_{k}^{}|m_{\omega}^{},(m_e^{})_l^{}|m_{\kappa}^{})\;,
\end{equation}
where $(m_e^{})_i$ are charged lepton masses,  and
 the induced mass matrix $(m_\nu^{})_{ij}^{}$ is defined in the effective
Lagrangian as
\begin{equation}
 \mathcal{L}_{\nu}^{} = - \sum_{i,j=1}^3 \frac{1}{2} (\overline{\nu}^c_L)_i^{} (m_\nu^{})_{ij} (\nu_L^{})_j^{} +
  {\rm h.c.}, 
\end{equation}
and $I(m_{11}^{},m_{12}^{}|m_{21}^{},m_{22}^{}|M)$ is the two-loop integral function
defined as
\begin{align}
&I(m_{11}^{},m_{12}^{}|m_{21}^{},m_{22}^{}|M)
\nonumber\\
&= \frac{1}{\pi^4}
\int d^4 p
\int d^4 q
\frac{1}{(p^2+m_{11}^2)}
\frac{1}{(p^2+m_{12}^2)}
\frac{1}{(q^2+m_{21}^2)}
\frac{1}{(q^2+m_{22}^2)}
\frac{M^2}{((p+q)^2+M^2)}\;.
\label{eq:I-func}
\end{align}
Following Refs.~\cite{vanderBij:1983bw},
one can evaluate the function $I(m_{11}^{},m_{12}^{}|m_{21}^{},m_{22}^{}|M)$ as
\begin{align}
&I(m_{11}^{},m_{12}^{}|m_{21}^{},m_{22}^{}|M)\nonumber\\
&=\frac{M^4\left\{I(m_{12}^{}|m_{22}^{}|M)-I(m_{11}^{}|m_{22}^{}|M)
	-I(m_{12}^{}|m_{21}^{}|M)+I(m_{11}^{}|m_{21}^{}|M)\right\}}
{(m_{11}^2-m_{12}^2)(m_{21}^2-m_{22}^2)}\;,
\end{align}
where
\begin{equation}
I(m_1^{}|m_2^{}|M)=-\left\{
	\frac{m_1^2}{M^2}f\left(\frac{m_2^2}{m_1^2},\frac{M^2}{m_1^2}\right)
	+\frac{m_2^2}{M^2}f\left(\frac{m_1^2}{m_2^2},\frac{M^2}{m_2^2}\right)
	+f\left(\frac{m_1^2}{M^2},\frac{m_2^2}{M^2}\right)
	\right\}\;.
\end{equation}
The function $f(x,y)$ is give by
\begin{align}
f(x,y)=&-\frac{1}{2}\ln x\ln y -\frac{1}{2}\left(\frac{x+y-1}{D}\right)\nonumber\\
&\times\left\{
	\mathrm{Li}_2^{}\left(\frac{-\sigma_{-}}{\tau_{+}}\right)
	+\mathrm{Li}_2^{}\left(\frac{-\tau_{-}}{\sigma_{+}}\right)
	-\mathrm{Li}_2^{}\left(\frac{-\sigma_{+}}{\tau_{-}}\right)
	-\mathrm{Li}_2^{}\left(\frac{-\tau_{+}}{\sigma_{-}}\right)\right.\nonumber\\
	&\phantom{Space}\left.
	+\mathrm{Li}_2^{}\left(\frac{y-x}{\sigma_{-}}\right)
	+\mathrm{Li}_2^{}\left(\frac{x-y}{\tau_{-}}\right)
	-\mathrm{Li}_2^{}\left(\frac{y-x}{\sigma_{+}}\right)
	-\mathrm{Li}_2^{}\left(\frac{x-y}{\tau_{+}}\right)\right\}\;,
\end{align}
where $D$, $\sigma_{\pm}$ and $\tau_{\pm}$ are
\begin{align}
D=&\sqrt{1-2(x+y)+(x-y)^2}\;,\nonumber\\
\sigma_{+}=&\frac{1}{2}(1-x+y+D)\;,\quad
\tau_{+}=\frac{1}{2}(1+x-y+D)\;,\quad\\
\sigma_{-}=&\frac{1}{2}(1-x+y-D)\;,\quad
\tau_{-}=\frac{1}{2}(1+x-y-D)\;,\quad\nonumber
\end{align}
and $\text{Li}_2(x)$ is the dilogarithm function defined as 
\begin{equation}
\text{Li}_2^{}(x)=-\int_0^x\frac{\ln(1-t)}{t}dt\;.
\end{equation}
We note that
in the limit of $m_{12}^{}=m_{22}^{}=0$ and $m_{11}^{}=m_{21}^{}=m_{\omega}^{}$ the
above function $I(m_{11}^{},m_{12}^{}|m_{21}^{},m_{22}^{}|M)$ 
has the same form as the function 
given in Refs.~\cite{macesanu,aristizabal},
\begin{equation}
I(m_{\omega}^{},0|m_{\omega}^{},0|m_{\kappa}^{})=-\int_0^1dx\int_0^{1-x}dy\frac{r}{x+(r-1)y+y^2}\ln\frac{y(1-y)}{x+ry}\;,
\end{equation}
where $r=m_{\kappa}^2/m_{\omega}^2$.
One can approximately estimate the above function as
\begin{equation}
I(m_{\omega}^{},0|m_{\omega}^{},0|m_{\kappa}^{})\sim 
\begin{cases}
%\left(0.5\ln^2\left(\frac{m_{\omega}^2}{m_{\kappa}^2}\right)
%+2.0\left(\frac{m_{\omega}^2}{m_{\kappa}^2}\right)^{-0.45}\right)\;,
2.8r\left(r+0.31\right)^{-1.5}\;,&(r\gtrsim 1)\\
%\frac{2.8m_{\kappa}^2}{m_{\omega}^2}\left(\frac{m_{\kappa}^2}{m_{\omega}^2}+0.31\right)^{-1.5}\;,
%& (m_{\omega}\lesssim m_{\kappa})\\
1.98r\left(r+0.12\right)^{-0.23}\;,&(r < 1)\\
%\frac{1.98m_{\kappa}^2}{m_{\omega}^2}\left(\frac{m_{\kappa}^2}{m_{\omega}^2}+0.12\right)^{-0.23}\;,
%& (m_{\omega} > m_{\kappa})\\
\end{cases}\;.
\end{equation}
Details of the Zee-Babu model have been studied 
in the literature\cite{macesanu,aristizabal,nebot,ohlsson}. 
It is known that the model can reproduce the present neutrino
data with satisfying constraints from the LFV.
 
We turn to the SUSY extension of the Zee-Babu model. 
The $\text{SU}(2)_{\text{L}}^{}$ singlet chiral superfields 
$\Omega_R^c$,
$\Omega_L^{}$,
$K_L^{}$,
and $K_R^c$
are added to the superfields in 
the minimal supersymmetric standard model (MSSM),
whose details are shown in 
Table.~\ref{mattercontent}.
Notice that although the non-SUSY Zee-Babu model includes only two  
$\text{SU}(2)_{\text{L}}^{}$ singlet scalars
these four chiral fields are required in the SUSY model.
If only $\Omega_R^c$ and $K_L^{}$ are introduced in the model, 
their fermion components are massless and the model is ruled out.
By introducing additional fields $\Omega_L^{}$ and $K_R^c$ such massless
fermions can be massive, and furthermore the model becomes anomaly free.
\begin{table}
\caption{Particle properties of relevant chiral superfields.}
\label{mattercontent}
\begin{center}
\begin{tabular}{|c||c|c||c|c|c||c|c|}\hline
&Spin 0&Spin 1/2&$\text{SU(3)}_{\text{C}}^{}$&$\text{SU(2)}_{\text{L}}^{}$&
$\text{U(1)}_{\text{Y}}^{}$&Electric charge&Lepton number\\ \hline
$L_i^{}$&
$\tilde{\ell}_{Li}^{}=\begin{pmatrix}\tilde{\nu}_{Li}^{}\\ \tilde{e}_{Li}^{}\end{pmatrix}$
&$\ell^{}_{Li}=\begin{pmatrix}\nu^{}_{Li}\\ e_{Li}^{}\end{pmatrix}
$&1&2&$-\frac{1}{2}$&$\begin{pmatrix}0\\ -1\end{pmatrix}$&1\\ \hline
$E_i^c$&$\tilde{e}_{Ri}^{*}$&$(e_R)^c$&1&1&1&1&$-1$\\ \hline
$\Phi_d^{}$
&$\phi_d^{}=\begin{pmatrix} \phi_d^0\\ \phi_d^-\end{pmatrix}$
&$\tilde{h}_d^{}=\begin{pmatrix} \tilde{h}_d^0\\ \tilde{h}_d^-\end{pmatrix}
$
&1&2&$-\frac{1}{2}$&$\begin{pmatrix} 0 \\ -1\end{pmatrix}$&0\\ \hline
$\Phi_u^{}$
&$\phi_u^{}=\begin{pmatrix}\phi_u^+\\ \phi_u^0\end{pmatrix}$
&$\tilde{h}_u=\begin{pmatrix}\tilde{h}_u^+\\ \tilde{h}_u^0\end{pmatrix}$
&1&2&$\frac{1}{2}$
&$\begin{pmatrix}1\\ 0 \end{pmatrix}$&0\\ \hline
$\Omega_R^c$&$\omega_R^*$&$(\tilde{\omega}_R)^c$&1&1&1&1&$-2$\\ \hline
$\Omega_L^{}$&$\omega_L^{}$&$\tilde{\omega}_L^{}$&1&1&$-1$&$-1$&2\\ \hline
$K_L^{}$&$\kappa_L^{}$&$\tilde{\kappa}_L^{}$&1&1&$-2$&$-2$&2\\ \hline
$K_R^c$&$\kappa_R^*$&$(\tilde{\kappa}_R)^c$&1&1&2&2&$-2$\\ \hline
\end{tabular}
\end{center}
\end{table}

The superpotential is given by\footnote{Hereafter we  omit
the summation symbol for simplicity.}
\begin{align}
W=&W_{\text{MSSM}}^{}+f_{ij}^{}L_i^{}\cdot L_j^{}\Omega_R^c+g_{ij}^{}E_i^cE_j^cK_L^{}
+\lambda_L^{}K_L^{}\Omega_R^c\Omega_R^c
+\lambda_R^{}K_R^c\Omega_L^{}\Omega_L^{}
\nonumber\\
&
+\mu_{\Omega}^{}\Omega_R^c\Omega_L^{}
+\mu_{K}^{}K_L^{}K_R^c\;,
\label{eq:W}
\end{align}
where $W_{\text{MSSM}}$ is the superpotential in the MSSM. 
The superfields in the superpotential are listed in Table.~\ref{mattercontent},
and the coupling matrices $f_{ij}$ and $g_{ij}$ are an
antisymmetric matrix $f_{ji}=-f_{ij}$  and a symmetric one $g_{ji}=g_{ij}$, respectively.
It is emphasised that 
we here impose the exact R-parity in order to protect the decay of the 
LSP,
so that the LSP is a candidate of the DM.
The soft SUSY breaking terms are given by 
\begin{equation}
\mathcal{L}_{\text{soft}}^{}=
\mathcal{L}_{\text{MSSM}}^{}
+\mathcal{L}_{\text{SZB}}^{}
+\mathcal{L}_{\text{C}}^{}\;,
\end{equation}
where $\mathcal{L}_{\text{MSSM}}$ represents the corresponding terms in the MSSM, 
\begin{align}
\mathcal{L}_{\text{SZB}}^{}=&
-M_+^2\omega_R^*\omega_R^{}
-M_-^2\omega_L^*\omega_L^{}
-M_{--}^2\kappa_L^*\kappa_L^{}
-M_{++}^2\kappa_R^*\kappa_R^{}
\nonumber\\
&
+\biggl(
-m_S^{}\tilde{f}_{ij}^{}\omega_R^*\tilde{\ell}_{Li}^{}\cdot \tilde{\ell}_{Lj}^{}
-m_S^{}\tilde{g}_{ij}^{}\kappa_L\tilde{e}_{Ri}^*\tilde{e}_{Rj}^*
-m_S^{}\tilde{\lambda}_L^{}\kappa_L^{}\omega_R^*\omega_R^*
-m_S^{}\tilde{\lambda}_R^{}\kappa_R^*\omega_L^{}\omega_L^{}
\nonumber\\
&\phantom{Spac}
-B_{\omega}^{}\mu_{\Omega}^{}\omega_R^*\omega_L^{}
-B_{\kappa}^{}\mu_{K}^{}\kappa_L^{}\kappa_R^*
+\text{h.c.}\biggr)
\;,
\label{eq:Lsoft}
\end{align}
and
\begin{align}
\mathcal{L}_{\text{C}}^{}=&
-C_u^{}\omega_R^*\phi_u^{\dagger}\phi_d^{}
-C_d^{}\omega_L^{}\phi_d^{\dagger}\phi_u^{}
-(C_{\omega}^{})^{ij}\omega_L^*\tilde{\ell}_{Li}^{}\cdot \tilde{\ell}_{Lj}^{}
+\text{h.c.}\;,
\label{eq:LsoftNH}
\end{align}
where $m_S^{}$ denotes a typical SUSY mass scale,
and $\tilde{f}_{ji}^{}=-\tilde{f}_{ij}^{}$ and $\tilde{g}_{ji}^{}=\tilde{g}_{ij}^{}$.
%Here $\omega_R^*$, $\omega_L$, $\kappa_L$, $\kappa_R^*$, $\tilde{l}_{Li}$, and $\tilde{e}_{Ri}^*$ are 
%the corresponding scalar components of the chiral superfields.
$\mathcal{L}_{\text{SZB}}^{}$ is the standard soft-breaking terms
with respect to the new charged singlet fields, $\omega_{L,R}^{}$ and 
$\kappa_{L,R}^{}$.
$\mathcal{L}_{\text{C}}^{}$  contains the terms so-called the ``C-terms''\cite{Hall:1990ac}, where the scalar component and its
conjugation are mixed
\footnote{The singlet scalar C-terms 
break SUSY hard, while the
terms listed in the $\mathcal{L}_{\text{C}}^{}$ include non-singlet scalars
and the quadratic divergence does not occur.}.

\begin{figure}
\begin{center}
{\includegraphics[scale=0.8]{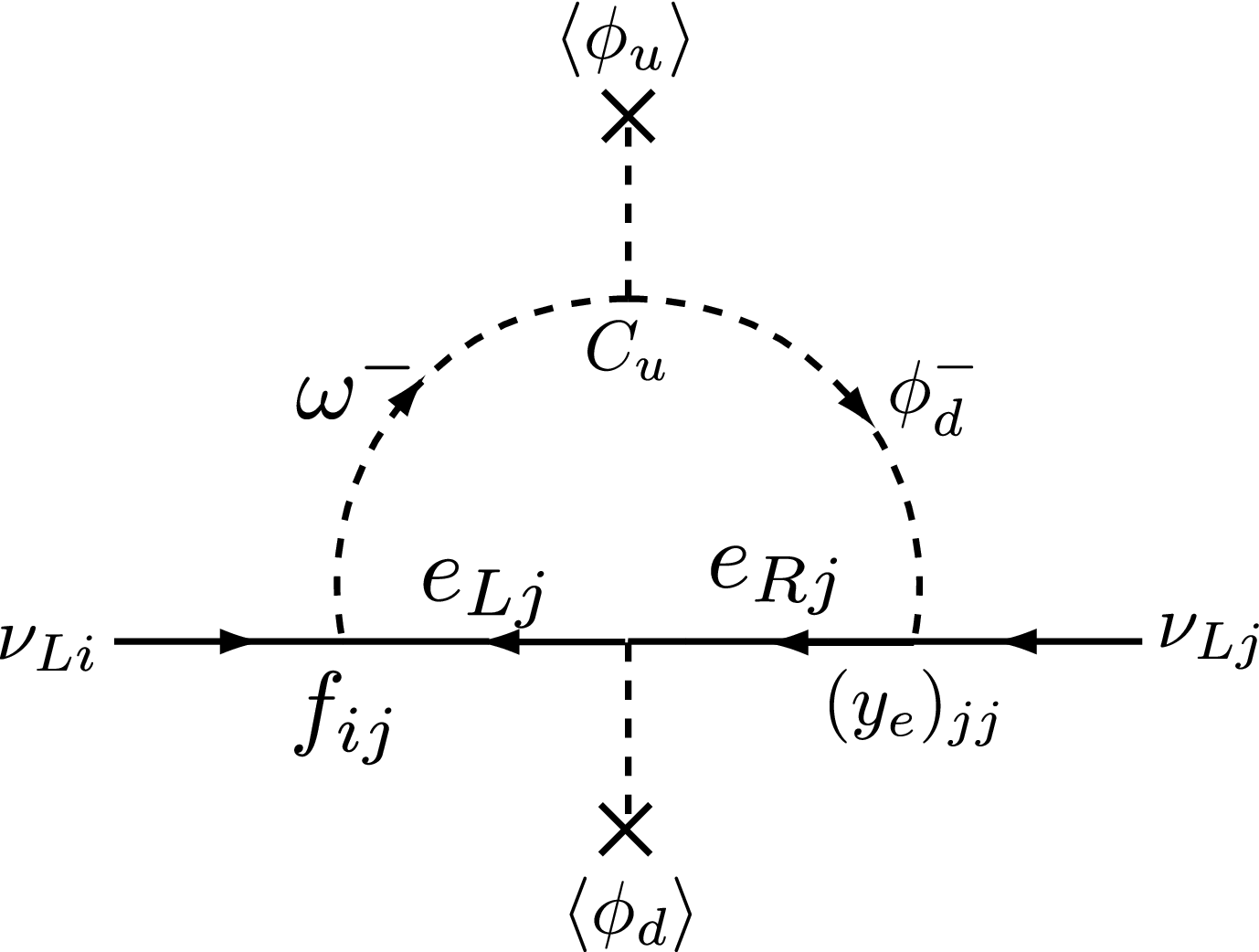}}
\end{center}
\caption{The one-loop diagram relevant to the neutrino mass matrix
 with the C-term $\omega^+\phi_u^*\phi_d^{}$. $(y_e^{})_{jj}^{}$ is 
 the charged lepton Yukawa coupling.}
\label{fig:diagram00}
\end{figure}

There are two possibilities in building a SUSY model with the charged
singlet fields, depending on whether or not the C-terms are 
switched on in a SUSY breaking scenario\footnote{
Many models derived by $N=1$ supergravity do not lead to the C-terms
and if they are absent at the cut off scale, they do not appear through the radiative corrections\cite{JackJonesKord}.
Thus the C-terms are usually ignored in the MSSM.
On the other hand, it is known that C-terms are induced in some models of SUSY breaking 
such as an intersecting brane model with a flux compactification\cite{Camara:2003ku}.}.
If we assume that $\mathcal{L}_{\text{C}}^{}$ is absent,
tiny neutrino masses are generated only by at least two loop diagrams as 
in the Zee-Babu model.
On the other hand, 
with the term $\omega_R^*\phi_u^{\dagger}\phi_d^{}$, tiny
neutrino masses are dominated by 
one loop diagrams in Fig.~\ref{fig:diagram00} just like in 
the original Zee model\cite{zee}.
In this Letter we focus on the case where the SUSY breaking mechanism does not lead to the soft SUSY breaking 
C-terms, so that all the neutrino masses are generated at the two loop
level.  
  The case with the C-term will be discussed elsewhere\cite{tmp}.  

%Let us also comment on the R-parity. We assume that the R-parity is conserved in the model. If small R-parity violation is 
%introduced, one loop diagrams with the right-handed sleptons can dominate the neutrino mass generation again.
%In this case, the two loop contribution of the new fields is sub-dominant even if the non-holomorphic terms are absent.

From the superfields $\Omega_R^c$, $\Omega_L^{}$, $K_L^{}$ and 
$K_R^c$, there appear singly charged ($Y=-1$)
and doubly charged ($Y=-2$) singlet scalar bosons,
$\omega_{R,L}^{}$ and $\kappa_{L,R}^{}$,  
as well as their superpartner fermions,
namely singly and doubly charged singlinos, 
$\tilde{\omega}$ and $\tilde{\kappa}$, respectively.
The superpotential and the soft SUSY breaking terms lead to 
the mass matrix for the singly charged scalars in the basis of
$(\omega_R^{},\omega_L^{})$ as\;,
\begin{equation}
M_{\omega}^2=
\begin{pmatrix}
M_+^2+|\mu_{\Omega}^2|-m_W^2\tan^2\theta_W^{}\cos2\beta& B_{\omega}^{}\mu_{\Omega}^{}\\
B_{\omega}^{}\mu_{\Omega}^{}&M_-^2+|\mu_{\Omega}^2|+m_W^2\tan^2\theta_W^{}\cos2\beta
\end{pmatrix}\;,
\end{equation}
and the mass matrix for the doubly charged singlet scalars in the basis of 
$(\kappa_L^{}, \kappa_R^{})$ as
\begin{equation}
M_{\kappa}^2=
\begin{pmatrix}
M_{--}^2+|\mu_{K}^2|+2m_W^2\tan^2\theta_W^{}\cos2\beta& B_{\kappa}^{}\mu_{K}^{}\\
B_{\kappa}^{}\mu_{K}^{}&M_{++}^2+|\mu_{K}^2|-2m_W^2\tan^2\theta_W^{}\cos2\beta
\end{pmatrix}\;,
\end{equation}
where $\tan\beta$ is 
a ratio of the two vacuum expectation values of the MSSM Higgs bosons as
$\tan\beta = \langle \phi_u^{}\rangle/\langle \phi_d^{}\rangle$.
As easily seen from the above expressions, 
$\omega_R^{}$  and 
$\omega_L^{}$ ($\kappa_L^{}$ and $\kappa_R^{}$) can mix with each other
by the soft-breaking ``B-term'', $(B_{\omega}^{}\mu_{\Omega}^{}) \omega_R^*\omega_L^{}$ 
($(B_{\kappa}^{}\mu_{K}^{})\kappa_L^{}\kappa_R^*$).
The mass eigenvalues of singly and doubly charged singlet scalar bosons
are obtained after diagonalising their mass matrices $M_{\omega}^2$ 
and $M_{\kappa}^2$ by the unitary matrices $U_{\omega}^{}$ and 
$U_{\kappa}^{}$ as
\begin{equation}
U_{\omega}^{\dagger}M_{\omega}^2U_{\omega}^{}
=\begin{pmatrix}
(m_{\omega})_1^2&0\\
0&(m_{\omega})_2^2
\end{pmatrix}\;,\quad
U_{\kappa}^{\dagger}M_{\kappa}^2U_{\kappa}^{}
=\begin{pmatrix}
(m_{\kappa})_1^2&0\\
0&(m_{\kappa})_2^2
\end{pmatrix}\;.
\end{equation}
The mass eigenstates are then given by 
\begin{equation}
\omega_{a}^{}=(U_{\omega}^{\dagger})_{a1}^{}\omega_R^{}
+(U_{\omega}^{\dagger})_{a2}^{}\omega_L^{}\;,\quad
\kappa_{a}=(U_{\kappa}^{\dagger})_{a1}^{}\kappa_L^{}
+(U_{\kappa}^{\dagger})_{a2}^{}\kappa_R^{}\;,\quad (a=1,2)\;.
\end{equation}
The mass eigenstates of the singlinos are
\begin{equation}
\tilde{\omega}=\begin{pmatrix} \tilde{\omega}_L^{}\\ \tilde{\omega}_R^{}\end{pmatrix}\;,\quad
\tilde{\kappa}=\begin{pmatrix}\tilde{\kappa}_L^{}\\ \tilde{\kappa}_R^{}\end{pmatrix}\;,
\end{equation}
whose mass eigenvalues are given by 
the SUSY invariant parameters as 
$m_{\tilde{\omega}}^{}=\mu_{\Omega}^{}$ and $m_{\tilde{\kappa}}^{}=\mu_K^{}$,
respectively.

\begin{figure}
\begin{center}
\begin{tabular}{ccc}
%\raisebox{2mm}{\includegraphics[scale=0.8]{diagramI.pdf}}\\
\raisebox{20mm}{(a)}&\phantom{space}&\includegraphics[scale=0.7]{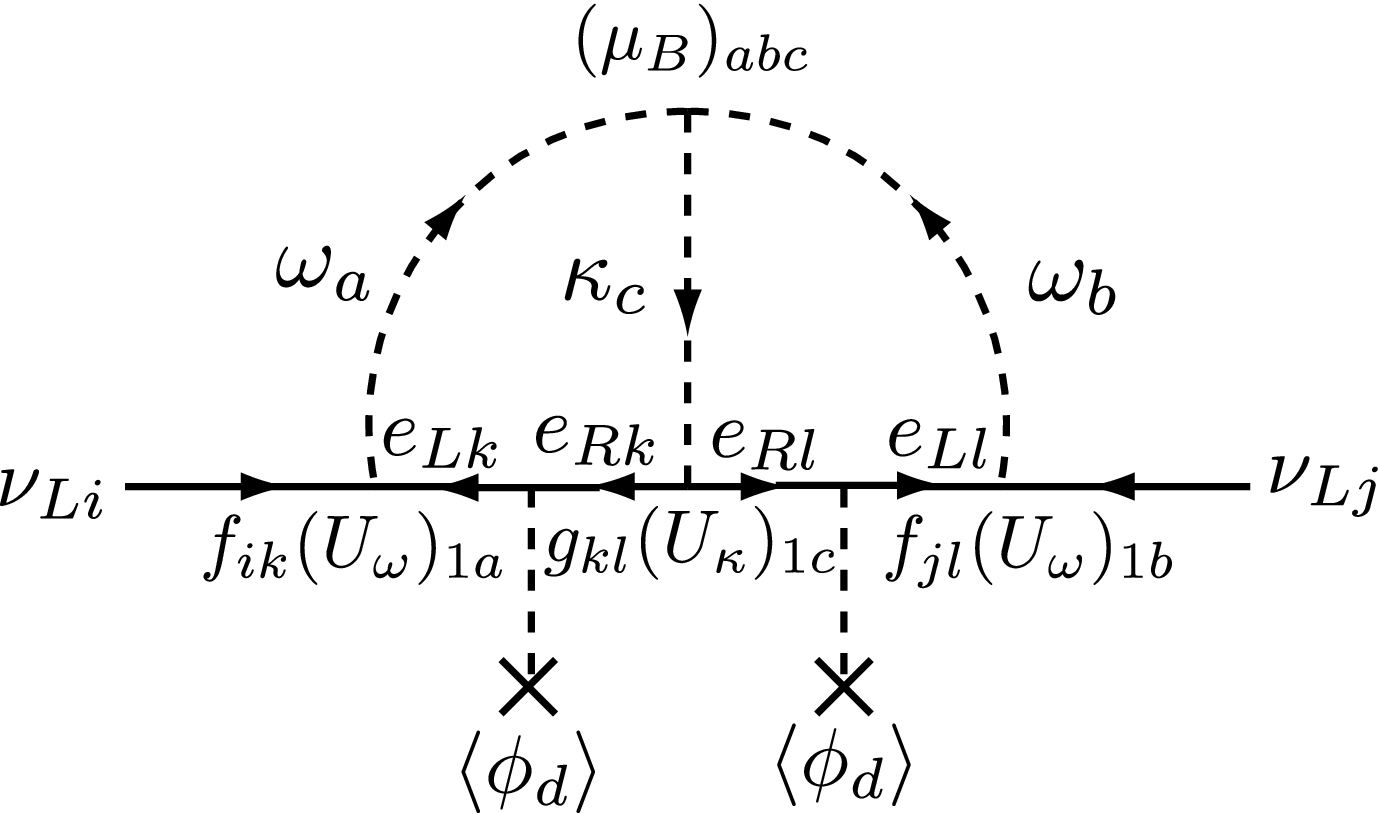}\\[2mm]
%\raisebox{9mm}{\includegraphics[scale=0.8]{diagramII.pdf}}\\
\raisebox{20mm}{(b)}&\phantom{space}&\includegraphics[scale=0.7]{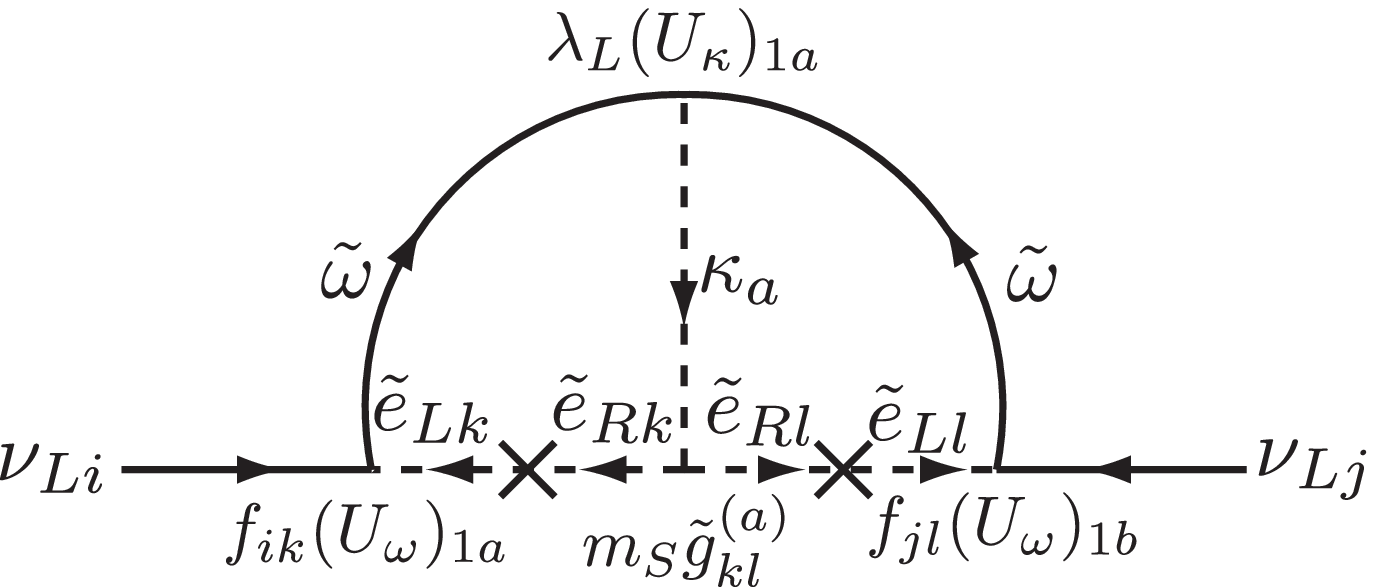}\\[2mm]
\raisebox{20mm}{(c)}&\phantom{space}&\includegraphics[scale=0.7]{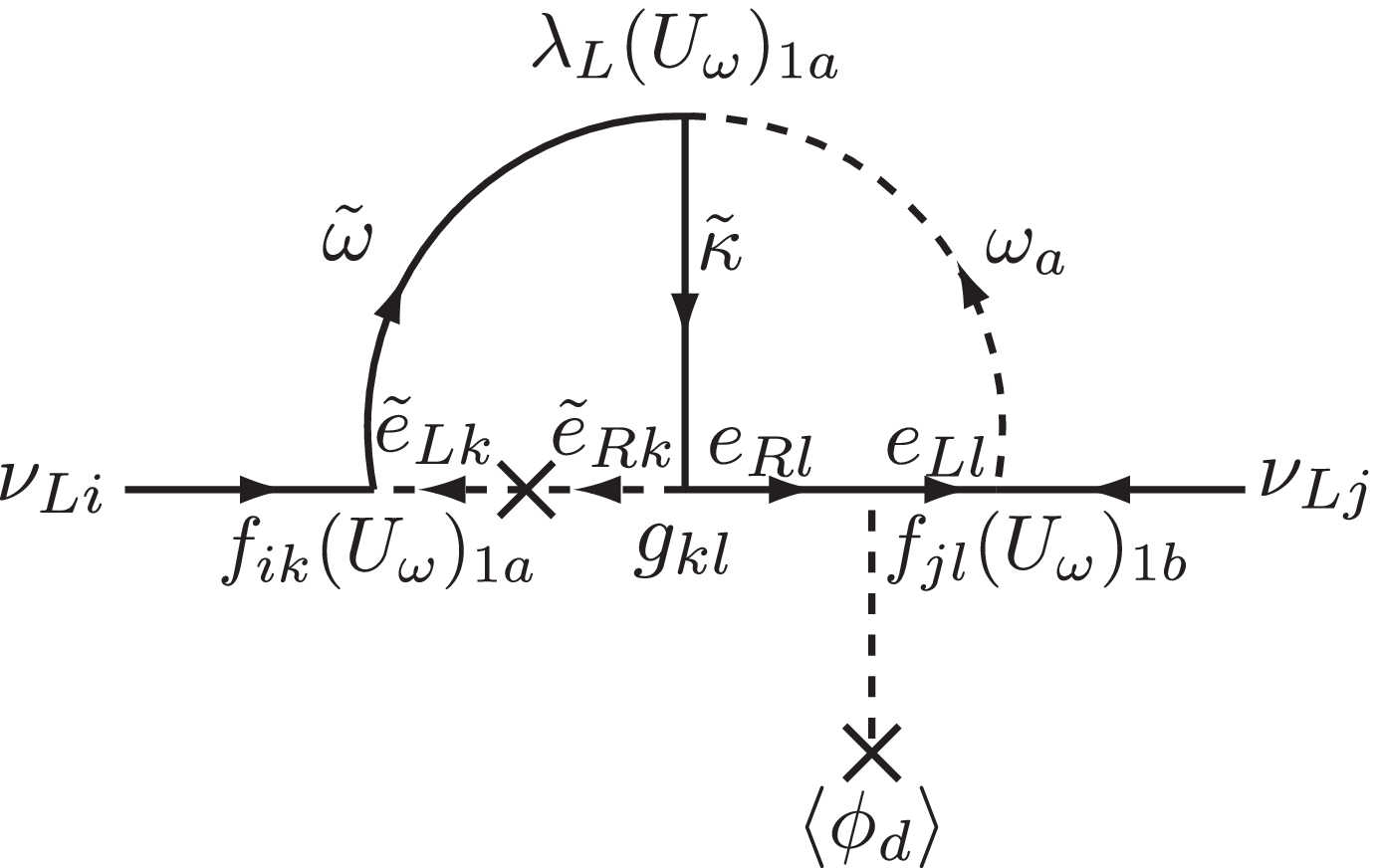}
\end{tabular}
\end{center}
\caption{The contributions to the neutrino mass generations. A type of a diagram (a) is 
the corresponding diagram to the non-SUSY Zee-Babu model. Diagrams (b) and (c) 
are new type of diagram in the SUSY model.}
\label{fig:diagram}
\end{figure}

The neutrino mass matrix is generated via 
the two-loop diagrams shown 
in Fig.~\ref{fig:diagram},
which can be written as
\begin{equation}
(m_{\nu}^{})_{ij}^{}=\frac{1}{(16\pi^2)^2}f_{ik}^{}(m_e)_{k}H_{kl}^{}(m_e)_{l}f_{jl}^{}\;,
\end{equation}
where the matrix $H_{kl}^{}$ is a symmetric matrix
\begin{align}
H_{kl}^{}=&
16(\mu_B^{})_{abc}^{}(U_{\omega}^{})^*_{1a}(U_{\omega}^{})^*_{1b} (U_{\kappa}^{})_{1c}^{}
g_{kl}^{} I((m_e^{})_{k}^{},(m_{\omega}^{})_a^{}|(m_e^{})_{l}^{},(m_{\omega}^{})_b^{}|(m_{\kappa})_c)
\nonumber\\
&
+16\frac{\lambda_{L}^*m_{\tilde{\omega}}^2}{m_S}(U_{\kappa}^{})^*_{1a}(U_{\kappa}^{})_{1a}^{}
\biggl\{
\frac{X_k^{}}{m_S^{}}
\tilde{g}_{kl}^{}
\frac{X_l^{}}{m_S^{}}
I((m_{\tilde{e}_R^{}}^{})_{k}^{},m_{\tilde{\omega}}^{}|(m_{\tilde{e}_{R}^{}})_l^{},m_{\tilde{\omega}}^{}|(m_{\kappa}^{})_a^{})
\nonumber\\
&\phantom{S}
+\frac{X_k^{}}{m_S^{}}
g_{kl}^{}
I((m_{\tilde{e}_{R}^{}})_k^{},m_{\tilde{\omega}}^{}|(m_{\tilde{e}_{L}^{}}^{})_l^{},m_{\tilde{\omega}}^{}|(m_{\kappa}^{})_a^{})
+g_{kl}
\frac{X_l^{}}{m_S^{}}
I((m_{\tilde{e}_{L}^{}}^{})_k^{},m_{\tilde{\omega}}^{}|(m_{\tilde{e}_{R}^{}}^{})_l^{},m_{\tilde{\omega}}^{}|(m_{\kappa}^{})_a^{})
\biggr\}
\nonumber\\
&
+\frac{8\lambda_{L}^{}m_{\tilde{\omega}}^{}m_{\tilde{\kappa}}^{}}{m_S^{}}(U_{\omega}^{})_{1a}^{}(U_{\omega}^{})_{1a}^*
\nonumber
\\
&\phantom{S}\times
\left\{
\frac{X_k^{}}{m_S^{}}
g_{kl}^{}
I((m_{\tilde{e}_{R}^{}}^{})_k^{},m_{\tilde{\omega}}^{}|(m_{e}^{})_l^{},(m_{\omega}^{})_a^{}|m_{\tilde{\kappa}}^{})
+
g_{kl}^{}
\frac{X_l^{}}{m_S^{}}
I((m_e^{})_{k}^{},(m_{\omega}^{})_a^{}|(m_{\tilde{e}_{R}^{}}^{})_l^{},m_{\tilde{\omega}}^{}|m_{\tilde{\kappa}}^{})
\right\}
\;,
\label{eq:Hkl}
\end{align}
where the indices $a,b,c$ run from 1 to 2, the mass eigenstates of the 
charged singlet scalars, 
$(m_{\tilde{e}_{R}^{}}^{})_i$ and $(m_{\tilde{e}_{L}^{}}^{})_i^{}$ are 
slepton masses,
the left-right mixing term in the slepton sector is parameterized as
$(m_{e}^{})_k^{}X_k^{}/m_S^{}$,
$I(m_{11}^{},m_{12}^{}|m_{21}^{},m_{22}^{}|M)$ is the loop function given in Eq.~(\ref{eq:I-func}), and
the other parameters are defined in the 
relevant Lagrangian as
\begin{align}
\mathcal{L}=&-2f_{ij}^{}(U_{\omega}^{})^*_{1a}\bar{\nu}_i^cP_L^{}e_j^{}\omega_a^*
-g_{ij}^{}(U_{\kappa}^{})_{1a}^{}\bar{e}_i^{}P_L^{}e_j^c\kappa_a^{}
-2f_{ij}^{}\tilde{\nu}_{Li}^*\bar{\tilde{\omega}}P_L^{}e_j^{}
-2f_{ij}^{}\bar{\nu}_i^cP_L^{}\tilde{\omega}^c\tilde{e}_{Lj}^{}
\nonumber\\
&
-2g_{ij}^{}\tilde{e}_{Ri}^*\bar{e}_j^{}P_L^{}\tilde{\kappa}
-\lambda_L^{}(U_{\kappa}^{})_{1a}^{}\bar{\tilde{\omega}}P_L^{}\tilde{\omega}^c\kappa_a^{}
-2 \lambda_L^{}(U_{\omega}^{})_{1a}^*\bar{\tilde{\omega}}P_L^{}\tilde{\kappa}\omega_a^*
-g_{ij}^{}(U_{\kappa}^{})_{1a}^{}(m_{e}^{})_j^{}\tilde{e}_{Ri}^*\tilde{e}_{Lj}^*\kappa_a^{}
\nonumber\\
&
-(\mu_B^{})_{abc}\omega_a^{}\omega_b^{}\kappa_c^*
-m_S^{}\tilde{g}_{ij}(U_{\kappa}^{})_{1a}\tilde{e}_{Ri}^*\tilde{e}_{Rj}^*\kappa_a^{}
+\text{h.c.}\;,
\end{align}
with
\begin{align}
(\mu_B)_{abc}\equiv&
A_L^*(U_{\omega}^{})_{1a}^{}(U_{\omega}^{})_{1b}^{}(U_{\kappa}^{})_{1c}^*
+A_R^{}(U_{\omega}^{})_{2a}^{}(U_{\omega}^{})_{2b}^{}(U_{\kappa}^{})_{2c}^*\;.
\end{align}
In the above expression, we assume that there is no flavour mixing in the
slepton sector.
In our model, there are two sources of the LFV processes.
One is the slepton mixing which also appear in the MSSM.
The other is the flavour mixing in the coupling with 
the charged singlet particles.
In order to concentrate on the latter contribution to the 
lepton flavour violating phenomena, the usual slepton mixing
effect is assumed to be zero.
The phenomenological constraints in our discussion strongly 
depend on this assumption.
If the assumption is relaxed, the phenomenological allowed 
parameters of the model can be changed to some extent.
Still we think our assumption is valuable to consider in order 
to obtain some definite physics consequences which are 
relevant to the new particles in our model.

%\section{Allowed parameter region under the current constraint}

It is non-trivial whether there is an allowed parameter region in our
model except for the decoupling limit where masses of all the super partner
particles are set to be much larger than the electroweak scale.
Let us search for the parameter region where the neutrino mixing is consistent with the 
present oscillation data and the LFV constraints are satisfied.

Flavour violation in couplings between $\text{SU}(2)_{\text{L}}^{}$ singlet fields and leptons
should be large in order to generate large off-diagonal elements in the neutrino mass matrix.
These large flavour violation couplings enhance the LFV processes.
In particular doubly charged singlet scalar exchange tree level diagram contributes to the 
$e_i^+ \to e_j^+e_k^+e_l^- $ process. 
The predicted decay width of $e_i^+\to e_j^+e_k^+e_l^-$ in the model is calculated as\cite{macesanu,aristizabal}
\begin{equation}
\Gamma(e_i^+\to e_j^+e_k^+e_l^-)=
C_{jk}^{}\frac{1}{8}\frac{(m_e^{})_i^5}{192\pi^3}\left|
(U_{\kappa}^{})_{1a}^*(U_{\kappa}^{})_{1a}^{}\frac{g_{il}^{}g_{jk}^{*}}{(m_{\kappa}^{})_a^2}\right|^2\;,
\end{equation}
where $C_{jk}$ is a statistical factor as
\begin{equation}
C_{jk}=\begin{cases}
1&(j=k)\\
2&(j\neq k)
\end{cases}\;.
\end{equation}
There can be still large contributions to 
$\mu\to e\gamma$,
even if the constraint from $\mu^+\to e^+e^+e^-$ can be avoided.
The contribution is from one-loop diagrams. The decay width of $e_i\to e_j\gamma$ is evaluated as
\begin{equation}
\Gamma(e_i\to e_j \gamma)=\frac{\alpha_e}{4}(m_e^{})_i^5\left(|A_L^{ji}|^2+|A_R^{ji}
|^2\right)\;,
\end{equation}
with 
\begin{align}
A_L^{ji}=&
-\frac{1}{(4\pi)^2}
\left\{
(U_{\omega}^{})_{1a}^*(U_{\omega}^{})_{1a}^{}
\frac{4f_{kj}^*f_{ki}^{}}{(m_{\omega}^{})_a^2}F_2\left(\frac{(m_{\nu}^{})_k^2}{(m_{\omega}^{})_a^2}\right)
-\frac{4f_{kj}^*f_{ki}^{}}{(m_{\tilde{\nu}_{L}^{}})_k^2}F_1\left(\frac{m_{\tilde{\omega}}^2}{(m
_{\tilde{\nu}_{L}^{}}^{})_k^2}\right)
\right\}
\;,\\
A_R^{ji}=&
-\frac{1}{(4\pi)^2}
\left\{
(U_{\kappa})_{1a}^*(U_{\kappa})_{1a}
\frac{g_{kj}^{*}g_{ki}^{}}{(m_{\kappa}^{})_a^2}\left(2F_2\left(\frac{(m_{e}^{})_k^2}{(m_{\kappa}^{})_a^2}\right)
+F_1\left(\frac{(m_{e}^{})_k^2}{(m_{\kappa}^{})_a^2}\right)\right)
\right.
\nonumber\\
&\phantom{SpaceSpace}
\left.-\frac{g_{kj}^{*}g_{ki}^{}}{(m_{\tilde{e}_{R}^{}}^{})_k^2}
\left(
2F_1\left(\frac{m_{\tilde{\kappa}}^2}{(m_{\tilde{e}_{R}^{}}^{})_k^2}\right)
+F_2\left(\frac{m_{\tilde{\kappa}}^2}{(m_{\tilde{e}_{R}^{}}^{})_k^2}\right)\right)
\right\}\;,
\end{align}
where $(m_{\nu})_i$ are neutrino masses, and $(m_{\tilde{\nu}_L^{}})_i$ are sneutrino masses.
The loop functions $F_1(x)$ and $F_2(x)$ are\cite{inamilim} 
\begin{align}
F_1(x)=&\frac{x^2-5x-2}{12(x-1)^3}+\frac{x\ln x}{2(x-1)^4}\;,\\
F_2(x)=&\frac{2x^2+5x-1}{12(x-1)^3}-\frac{x^2\ln x}{2(x-1)^4}\;.
\end{align}
The coupling constants $f_{ij}$ only have nonzero values in flavour off-diagonal
 elements, and they tend to be large to reproduce the bi-large mixing.
Then the bound from  the data becomes severe.

Let us discuss how the LFV processes constrain the parameter space.
First of all, the tree level diagram contributing to the $\mu\to eee$ must be 
suppressed.
The present bound on the branching fraction  is
$B(\mu^+\to e^+e^+e^-)<1.0\times 10^{-12}$\cite{Bellgardt:1987du},
which gives very strong constraint on the model parameter space.
There are two possible cases to suppress the tree level contribution to the $\mu^+\to e^+e^+e^-$.
The first possibility is considering heavy doubly charged bosons $\kappa_1$ and $\kappa_2$.
If $g_{11}\sim g_{12}\sim 0.1$ is taken, the doubly charged bosons 
should be heavier than 15 TeV to avoid too large contribution.
The second option is suppressing a product of the couplings $|g_{12}g_{11}|$.
When the doubly charged bosons are $500$\;GeV, 
the upper bound on the product $|g_{12}g_{11}|$ is obtained as $|g_{12}g_{11}|<10^{-5}$.
The contributions to $\tau^+\to e^+e^+e^-$, 
$\tau^+\to e^+e^+\mu^-$, 
$\tau^+\to \mu^+\mu^+e^-$,
$\tau^+\to \mu^+\mu^+\mu^-$,
$\tau^+\to \mu^+ e^+e^-$, 
and
$\tau^+\to e^+\mu^+\mu^-$  can be computed in the same manner.
These flavour changing tau decays into three leptons are also enhanced in the model
with tree level contributions.
If future tau flavour experiments such as the high luminosity B factories\cite{superB} would discover a signal of such decays,
it could support the model.
In the phenomenological point of view, the scenario with a light doubly charged singlet scalar
is attractive because the scenario with such a light exotic particle is
testable at the LHC.
Therefore we have searched for a solution with a suppressed $|g_{12}g_{11}|$ and we have found that
the coupling $g_{11}$ can be taken to be so small that the tree level contribution 
to the $\mu^+ \to e^+e^+e^-$ process is negligible with reproducing the neutrino oscillation data.
In such a parameter space, the $B(\mu^+\to e^+e^+e^-)$ is suppressed by the electromagnetic coupling constant 
compared with $B(\mu\to e\gamma)$,
say $B(\mu^+\to e^+e^+e^-)\sim \alpha_eB(\mu\to e\gamma)$ where
the current upper limit is given by $B(\mu\to e\gamma) < 1.2\times
10^{-11}$ \cite{muegamma}.
The $B(\mu^+\to e^+e^+e^-)$ is below the experimental upper bound, if the constraint of $B(\mu\to e\gamma)$
is satisfied. 
%The $\mu\to e\gamma$ constraint gives the strongest constraint among the LFV processes.

In our analysis below, we work in the limit of 
$B_{\omega}^{}\mu_{\Omega}^{}\to 0$ and $B_{\kappa}^{}\mu_K^{}\to 0$ for simplicity.
If these terms are switched on, the mixings in the charged singlet 
scalar mass eigenstates take part in the neutrino mass generation.
However these mixings do not change our main results.
In this limit, the mixing matrices $U_{\omega}^{}$ and $U_{\kappa}^{}$ 
become  the unit matrix, and only $\omega_1^{}$ and $\kappa_1^{}$ 
contribute to the neutrino mass matrix and the LFV.
Below we simply write the relevant fields as 
$\omega\equiv\omega_1^{}$ and $\kappa\equiv\kappa_1^{}$,
and their masses are written as 
$m_{\omega}^{}\equiv (m_{\omega}^{})_1$ and 
$m_{\kappa}^{}\equiv (m_{\kappa}^{})_1$.

Following the above strategy, we search for an allowed 
parameter set.
An example of the allowed parameter sets is 
\begin{align}
&
f_{12}^{}=f_{13}^{}=\frac{f_{23}^{}}{2}=3.7\times 10^{-2}\;,\quad
\nonumber\\
&g_{11}^{}\simeq 0\;,\quad g_{12}^{}=4.8\times 10^{-7}\;,\quad
g_{13}^{}=2.1\times 10^{-7}\;,\quad
\nonumber\\
&
g_{22}^{}=-0.13\;,\quad
g_{23}^{}=6.1\times 10^{-3}\;,\quad
g_{33}^{}=-4.6\times 10^{-4}\;,\quad
\nonumber\\
&
\tilde{g}_{ij}^{}=g_{ij}^{}\;,\quad
\lambda_a^{}=1.0\;,\quad
\mu_B^{}=500\;\text{GeV}\;,\quad
\frac{X_k^{}}{m_S^{}}=1.0\;,\nonumber\\
&(m_{\tilde{e}_{L}^{}}^{})_k=(m_{\tilde{e}_{R}^{}}^{})_k=(m_{\tilde{\nu}_L^{}}^{})_k^{}=m_S^{}=1000\;\text{GeV}\;,\quad
\nonumber\\
&
m_{\omega}^{}=600\;\text{GeV}\;,\quad
m_{\tilde{\omega}}^{}=600\;\text{GeV}\;,\quad
m_{\kappa}^{}=300\;\text{GeV}\;,\quad
m_{\tilde{\kappa}}^{}=200\;\text{GeV}\;,\quad
\nonumber\\
&(m_{\omega}^{})_2^{}\gg m_{\omega}\;,\quad
(m_{\kappa}^{})_2^{}\gg m_{\kappa}\;.
\label{eq:benchmark}
\end{align}
On this benchmark point, the neutrino masses and mixing angles are given as
\begin{align}
&\sin^2\theta_{12}=0.33\;,\quad 
\sin^2\theta_{23}=0.5\;,\quad
\sin^2\theta_{13}=0.0\;,\nonumber\\
&\Delta m_{21}^2=7.6\times 10^{-5}\;\text{eV}^2\;,\quad
|\Delta m_{31}^2|=2.5\times 10^{-3}\;\text{eV}^2\;,
\end{align}
which are completely consistent with the present neutrino data:
the global data analysis\cite{Schwetz:2008er} of the neutrino oscillation experiments provide 
$\sin^2\theta_{12}=0.318^{+0.019}_{-0.016}$, 
$\sin^2\theta_{23}=0.50^{+0.07}_{-0.06}$, 
$\sin^2\theta_{13}=0.013^{+0.013}_{-0.009}$, 
$\Delta m_{21}^2=(7.59^{+0.23}_{-0.18})\times 10^{-5}\;\text{eV}^2$,
and
$|\Delta m_{31}^2|=(2.40^{+0.12}_{-0.11})\times 10^{-3}\;\text{eV}^2$.
Based on this benchmark point, our model predicts
$B(\mu\to e\gamma)=1.1\times 10^{-11}$ 
and $B(\tau^+\to  \mu^+ \mu^+ \mu^-)=1.3\times 10^{-8}$, 
both of which are just below the present experimental bounds.
Since the LFV is naturally enhanced in the model, the MEG
experiment\cite{MEG}, 
which is expected to achieve $B(\mu\to e\gamma)<10^{-13}$ in a few years,
will cover very wide regions of the parameter space.
Apart from the bench mark scenario, there can be other parameter sets
where the neutrino data and LFV data are satisfied.  However, we here
do not discuss details for such a possibility.  A more general survey of the
parameter regions may be performed elsewhere\cite{tmp}.

%\section{Phenomenology}
We turn to discuss collider phenomenology in  the model assuming
the parameters of the benchmark scenario given in Eq.~(\ref{eq:benchmark}).
In our model, the new $\text{SU}(2)_{\text{L}}^{}$ charged singlet fields
are introduced, which can be accessible at collider experiments such
as the LHC unless they are too heavy.
In particular, the existence of the doubly charged singlet scalar
boson and its SUSY partner fermion (the doubly charged singlino)
provides discriminative phenomenological signals.
They are produced in pair ($\kappa^{++}\kappa^{--}$ or
$\tilde{\kappa}^{++}\tilde{\kappa}^{--}$) and 
each doubly charged boson (fermion) can be observed as 
a same-sign dilepton event, 
which would be a clear signature.
In this Letter, we focus on such events including doubly charged particles.
%
%The decay modes of the doubly charged scalar bosons can be 
%$\kappa_{1,2}\to e_ie_j$, 
%$\kappa_{1,2}\to \omega_a\omega_b$,
%$\kappa_{1,2}\to \tilde{e}_{Ri}\tilde{e}_{Rj}$,
%$\kappa_{1,2}\to \tilde{\omega}\tilde{\omega}$ and 
%$\kappa_{1,2}\to \tilde{\chi}^0\tilde{\kappa}$,
%depending on the mass spectrum.
For the benchmark point given in Eq.~(\ref{eq:benchmark}),
almost all the $\kappa$ decays into the same-sign muon pair,
$\kappa^{\pm\pm}\to \mu^{\pm}\mu^{\pm}$.

\begin{figure}
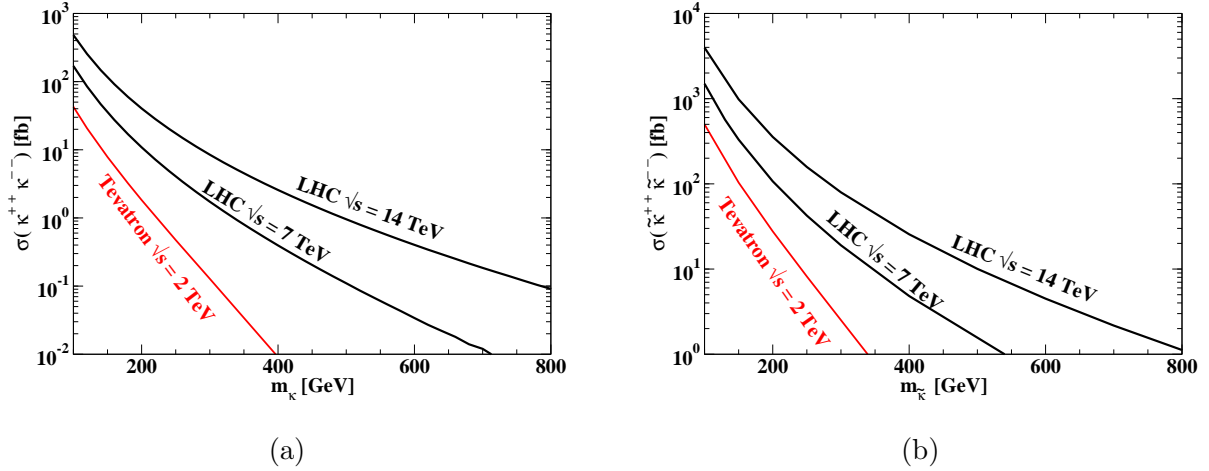

\begin{center}
\begin{tabular}{ccc}
\includegraphics[scale=0.3]{xs_k.eps}&\phantom{spac}
&\includegraphics[scale=0.3]{xs_kino.eps}\\
(a)&&(b) %\\ \\
%\includegraphics[scale=0.3]{xs_w.eps}&
%\includegraphics[scale=0.3]{xs_wino.eps}\\
%(c)&(d)
\end{tabular}
\end{center}
\caption{Production cross sections of 
 (a) $\kappa^{++}\kappa^{--}$ and   
 (b) $\tilde{\kappa}^{++}\tilde{\kappa}^{--}$,
% (c) $\omega_a^{+}\omega_a^{-}$,  and 
% (d) $\tilde{\omega}^{+}\tilde{\omega}^{-}$ 
via Drell-Yan processes at the LHC ($pp$)
 and the Tevatron ($p\overline{p}$).
 The production cross section at the LHC is evaluated 
for $\sqrt{s}=7\;\text{TeV}$ and $\sqrt{s}=14\;\text{TeV}$.
}
\label{fig:xs}
\end{figure}
At hadron colliders such as the LHC and the Tevatron,
the doubly charged singlet scalar $\kappa$ and the doubly charged singlino 
$\tilde{\kappa}$ are produced dominantly  in pair through 
the Drell-Yang processes. 
The production cross sections for 
$\kappa^{++}\kappa^{--}$ and 
$\tilde{\kappa}^{++}\tilde{\kappa}^{--}$ are shown as
in Fig.~\ref{fig:xs}(a) and Fig.~\ref{fig:xs}(b), respectively.
The first two plots from above correspond to the cross sections
at the LHC of $\sqrt{s}=14$ TeV and $\sqrt{s}=7$ TeV, and the
lowest one does to that at the Tevatron of $\sqrt{s}=2$ TeV.
We note that magnitudes of the production cross sections
for the pair of singly-charged singlet scalars  
$\omega^+\omega^-$ and that of singly-charged singlinos
$\tilde{\omega}^+\tilde{\omega}^-$ are (1/4) smaller than
those for $\kappa^{++}\kappa^{--}$ and 
$\tilde{\kappa}^{++}\tilde{\kappa}^{--}$ for the common mass
for produced particles.
The direct search of doubly charged Higgs bosons at the Tevatron gives
the lower bound on the mass assuming large branching ratio decaying to muon pairs
as $m_{\kappa}^{}\gtrsim 150\;\text{GeV}$ \cite{:2008iy}.
Such a bound on the mass of doubly charged singlinos is partly discussed in
Ref.~\cite{singlino++}.
At the LHC with $\sqrt{s}=7$ TeV with the integrated luminosity
${\cal L}$ of 1 fb$^{-1}$, about 100 of $\tilde{\kappa}^{++}\tilde{\kappa}^{--}$ pairs
can be produced when $m_{\tilde{\kappa}}^{} = 200$ GeV, while only
a couple of the $\kappa^{++}\kappa^{--}$ pair is expected for $m_\kappa^{} =300$ GeV.

\begin{figure}
\begin{center}
\includegraphics[scale=0.49]{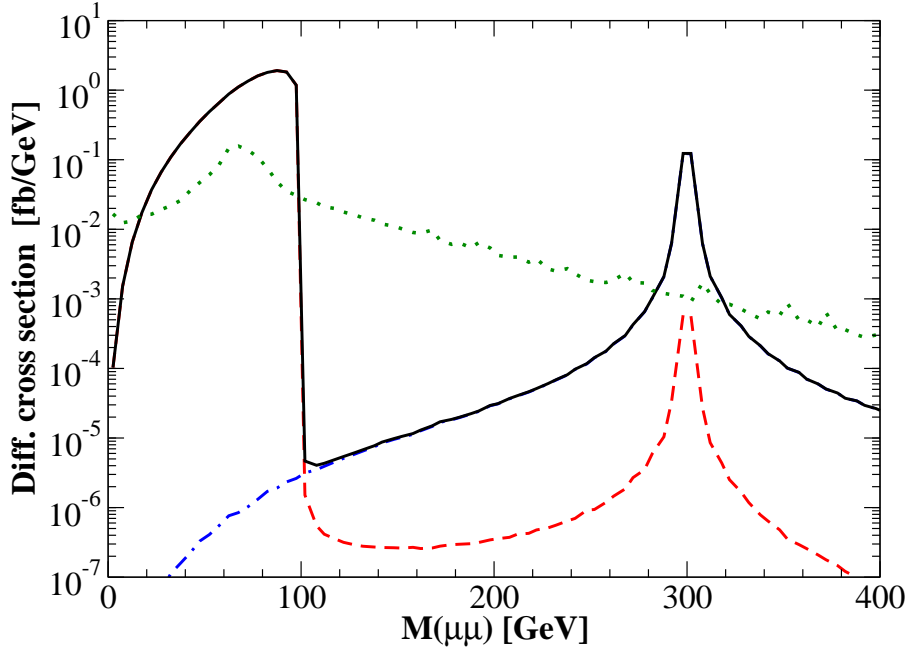}
\end{center}
\caption{
The invariant mass distribution of the same-sign dilepton event. 
The benchmark point in Eq.~(\ref{eq:benchmark}) is used
and the neutralino mass is taken as $m_{\tilde{\chi}^0}=100\;\text{GeV}$.
The dashed (red) curve corresponds to the events from 
$pp\to \tilde{\kappa}^{++} \tilde{\kappa}^{--}
   \to \tilde{\chi}^0 \kappa_1^{++} \tilde{\chi}^0 \kappa_1^{--} \to \tilde{\chi}^0
   \tilde{\chi}^0  \mu^+\mu^+\mu^-\mu^-$.
The dot-dashed (blue) curve shows the contributions from 
$pp\to \kappa^{++} \kappa^{--} \to \mu^+\mu^+\mu^-\mu^-$.
The solid (black) curve denotes total events from the both signal processes.
The dotted (green) curve shows the background events. 
For kinematical cut, see the text.
}
\label{fig:Mmm}
\end{figure}

In Fig.~\ref{fig:Mmm}, the distribution of the differential cross section for
four muon (plus a missing transverse momentum) final states as a function of 
the invariant mass $M(\mu^+\mu^+)$ of the same-sign muon pair is shown
assuming the bench mark scenario in Eq.~(\ref{eq:benchmark}) at the LHC with
$\sqrt{s}=7$ TeV.
In order to suppress background events, 
we select the muon events with 
the transverse momentum larger than 20\,GeV and 
the pseudo-rapidity less than 2.5.
The signal events come from both
$pp\to \kappa^{++} \kappa^{--} \to \mu^+\mu^+\mu^-\mu^-$ and
$pp\to \tilde{\kappa}^{++} \tilde{\kappa}^{--}
   \to \tilde{\chi}^0 \kappa_1^{++} \tilde{\chi}^0 \kappa_1^{--} \to \tilde{\chi}^0
   \tilde{\chi}^0  \mu^+\mu^+\mu^-\mu^-$.
The $M(\mu^+\mu^+)$ distribution can be a key to explore
the phenomena with the doubly charged particles. 
The doubly charged scalar mass and the mass difference between the
doubly charged singlino and the neutralino are simultaneously determined at the
LHC. A sharp peak is expected in the $M(\mu^+\mu^+)$ distribution at
$M(\mu^+\mu^+) = m_{\kappa}^{}$, because the same-sign muon pair 
 from the $\kappa$ decay is not associated with missing particles.
On the other hand, the doubly charged singlino decays as
$\tilde{\kappa}^{--}\to \tilde{\chi}^0\kappa^{--} \to \tilde{\chi}^0\mu^-\mu^-$ 
in the case that the lightest R-parity odd particle is a neutralino,
$\tilde{\chi}^0$, which is a DM candidate in the model.
In this Letter, we just assume that the LSP neutralino is Bino-like. 
In our analysis, we fix the neutralino mass as $m_{\tilde{\chi}^0}^{}=100\;\text{GeV}$.
The mass difference between $\tilde{\kappa}$ and $\tilde{\chi}^0$
can be measured by looking at a kink at
$M(\mu^+\mu^+)=m_{\tilde{\kappa}}^{}-m_{\tilde{\chi}^0}^{}$ in the $M(\mu^+\mu^+)$ distribution.
The main background comes from four muon events from the SM processes
where muons are produced via the $ZZ$, $\gamma\gamma$ and $\gamma Z$ production, 
or a pair production of muons with the $Z$ or $\gamma$ emission.
The expected background is also shown in Fig.~\ref{fig:Mmm}.
The events from signal dominate those from the background in the area
of $M(\mu^+\mu^+) < m_{\tilde{\kappa}}^{}-m_{\tilde{\chi}^0}^{}$ and
at around $M(\mu^+\mu^+) \sim m_{\kappa}^{}$.
The background events have been evaluated by using CalcHEP\cite{calchep}. 
From this rough evaluation, one may expect that the event from the
signal can be identified even at the LHC with $\sqrt{s}=7$ TeV
and ${\cal L}=1$ fb$^{-1}$.
As for the case with $\sqrt{s}=14\,\text{TeV}$, 
the signal to background ratio becomes
larger and it will be more promising to explore our model.

There are other models in which the same-sign dilepton events are predicted.
The model with the complex triplet scalar fields is an example of 
such a class of models\cite{triplet}.
They can in principle be distinguished by looking at the decay products
from doubly charged fields.
In our scenario, $\kappa^{\pm\pm}$ can mainly decay into  
$\mu^{\pm}\mu^{\pm}$, 
while in the triplet models where the decay of doubly charged singlet 
scalars are
directly connected with the neutrino mass matrix, there is no solution
where only the $\mu^{\pm}\mu^{\pm}$ mode can be dominant decay mode.
The difference in such decay pattern can be used to discriminate our model
from the triplet models.
%In the triplet Higgs model with the doubly charged scalar mass larger than $2m_W$, the doubly 
%charged scalar can decay into two same-sign $W$ bosons and this decay includes two missing neutrinos.
%Then the $M(\mu\mu)$ distribution differs from it in the SUSY Zee-Babu model****** CHECK !!! ******.
%It means that the $M(\mu\mu)$ distribution distinguishes these two models. 

%[PARAGRAPH OF OMEGA, OMEGA TILDE]
%[PARAGRAPH OF DM, BARYOGEN]
In this Letter, we have not discussed details for DM physics in our
model. Assuming the Bino-like LSP, we expect that our DM candidate
can satisfy the constraints from the WMAP data for the DM abundance
in a similar way to the case in the similar scenario in the MSSM. 
We still note that the existence of doubly and singly charged particles
in our model may change the cross section of DM pair
annihilation at the one loop level to some extent, so that they  
may affect the DM abundance. 
The detail is, however, beyond
the scope of this Letter, which will be discussed elsewhere\cite{tmp}. 

We also give a comment on the possibility for baryogenesis.
There can be several possibilities to realise baryogenesis in our model,
such as using the Affleck-Dine mechanism\cite{AD}, low energy leptogenesis\cite{ZBlepto}, and
electroweak baryogenesis (EWBG). 
In the scenario of EWBG, the electroweak phase transition must be of
strongly first order. 
In the MSSM the scenario of EWBG turns out to be rather challenging
\cite{ewbg-mssm2}.
On the contrary in our model, such a scenario may be natural and realistic.
In our model, there are doubly and singly charged singlet scalar fields.  
When they have non-decoupling property\cite{nondec,Aoki:2008av,aks_dm},
the parameter region of the strong first order phase transition can be
much wider than that in the MSSM. 
In addition, there can be many CP violating phases in the model, which
are also required for successful baryogenesis.

%\section{Conclusion}
We have discussed the SUSY extension of the Zee-Babu model
under R-parity conservation.   
In the model, it is not necessary to introduce very high energy
scale as compared to the TeV scale, and the model lies in the reach of 
the collider experiments
and the flavour measurements.
We have found that the neutrino data can be reproduced
with satisfying the current bounds from the LFV 
even in the scenario where not all the superpartner particles are heavy. 
The LSP can be a DM candidate.
Phenomenology of doubly charged singlet fields
has  also been discussed at the LHC.

The work of MA was supported in part by Grant-in-Aid for Young Scientists
(B) no. 22740137, 
that of SK was supported in part by Grant-in-Aid for Scientific Research
(A) no. 22244031 and (C) no. 19540277, and 
that of TS was supported in part by Grant-in-Aid for Scientific Research on Priority Areas,
no. 22011007.
%SK also acknowledges financial support from the PMI2
%Connect Initiative in the form of a Research Cooperation Grant.
The work of KY was supported by Japan Society for the Promotion of
Science (JSPS Fellow (DC2)). 

%%%%%%%%%%%%%%%%%%%%%%%%%%%%%%%%%

%%%%%%%%%%%%%%%%%%%%%%%%%%%%%%%%%
\end{document}